\documentclass[12pt]{article}
\newcommand{\myTr}{\ensuremath{\mathop{\rm Tr}\nolimits}}
\def\Tr(#1;#2;#3){\myTr_{#1}^{#2}(#3)}
\newcommand{\ket}[1]{\ensuremath{\left|#1\right\rangle}}
\newcommand{\QFT}{\ensuremath{\mathop{\rm QFT}\nolimits}}
\newcommand{\AQFT}{A}
\newcommand{\oneover}{\ensuremath{\frac{1}{\sqrt2}}}
\usepackage{amsmath}
\usepackage{epsfig}
\title{{\Large Approximation Errors}}
\author{Nathan W. Panike \\ {\sl National Security Agency, Suite 6409 } \\ {\sl 9800 Savage Road} \\ {\sl Fort George G. Meade, MD 20755}}
\begin{document}
\maketitle
\begin{abstract}
A bound on the error introduced by truncating a quantum addition is given.
This bound shows that only a few controlled rotation gates will be necessary to get a reliable computation.
\end{abstract}
\section*{Introduction}
Draper \cite{bib:Draper} introduced an addition algorithm for a quantum computer that uses only Hada\-mard and controlled rotation gates.
For an $n$-bit addition, his algorithm requires only $2n$ qubits, $O(n^2)$ gates in its full implementation, and allows for parallelization.
He also cited Barenco, et.\ al., to appeal that an approximate quantum Fourier transform may be more accurate than a full quantum Fourier transform \cite{bib:Barenco}.
In this paper, we investigate the error introduced by truncating the algorithm by removing all gates with tolerance less than a certain threshold, assuming that no decoherence is introduced into the computation.
This work shows that the probability of error decreases rapidly as one adds more gates, so that reliable computation can be achieved even with relatively few controlled rotation gates used in the approximate algorithm.
The reader is assumed to be familiar with Draper's paper, as this work will draw heavily from it as a source of ideas. 

From a high level, the full machinery of Draper's algorithm goes per the following diagram, where $\ket a$ and $\ket b$ are both $n$-bit quantum registers, and $\QFT_{2^n}$ is the quantum Fourier transform on $n$ qubits.
\begin{equation}
\label{eq:DraperHigh}
\begin{array}{ccccccc}
\ket b&\longrightarrow&\ket b&\longrightarrow&\ket b&\longrightarrow&\ket b\\
\ket a&\longrightarrow&\QFT_{2^n}\ket a&\longrightarrow&\QFT_{2^n}\ket{a+b}&\longrightarrow&\ket{a+b}\\
\end{array}
\end{equation}
Here $a+b$ represents the $n$-long bit string formed by the sum of $a$ and $b$ considered as numbers modulo $2^n$, not the vector sum of $\ket a$ and $\ket b$.

The $\QFT$ algorithm used in this computation is a refinement due to Coppersmith that shows how the $\QFT$ can be implemented using only Hadamard and controlled rotation gates \cite{bib:Coppersmith}.
A controlled rotation gate is given by the function $R_\theta$, which is the action
\[
R_\theta:\ket a\otimes\ket b\mapsto e^{i\theta ab}\ket a\otimes\ket b,
\]
where $a$ and $b$ are either 0 or 1.
We are supposing that we cannot perform a rotation $R_\theta$ for an angle \mbox{$|\theta|<2\pi/2^k$} for some $k$ because such a gate may cost too much to manufacture, may not be reliable, or the computation may be faster because of a fewer number of gates. 
The important number here is $k$ because we are calculating the error introduced by truncating a computation by eliminating all gates $R_\theta$, where \mbox{$|\theta|<2\pi/2^k$}.
We will call the maximal~$k$ the {\sl threshold} for our computation.
We would like to replace Diagram~\ref{eq:DraperHigh} with the following diagram, where $\AQFT$ is the approximate quantum Fourier transform on $n$ qubits with $k$ as the threshold, and $\Psi_b$ represents controlled rotations conditioned on $b$. 
\[
\begin{array}{ccccccc}
\ket b&\longrightarrow&\ket b&\longrightarrow&\ket b&\longrightarrow&\ket b\\
\ket a&\longrightarrow& \AQFT\ket a&\longrightarrow&\Psi_b\AQFT\ket a&\longrightarrow&\AQFT^\dag\Psi_b\AQFT\ket a\\
\end{array}
\]
We want to ensure that the quantity 
\[
\bigl|\langle a+b|\AQFT^\dag\Psi_b\AQFT|a\rangle\bigr|^2,
\]
which is the probability that the the approximate addition algorithm yields the same result as the full implementation, is close to unity for a given threshold~$k$.

\section*{Notation and Background}
A review of notation is needed at this point. 
We will follow Draper's notation by defining $e(t)=e^{2\pi it}$, so that $e(x+y)=e(x)e(y)$, and $e(x)=1$ if and only if $x$ is an integer.
We also define
\[
0.a_na_{n-1}\ldots a_1=\frac{a_n}{2}+\frac{a_{n-1}}{2^2}+\cdots+\frac{a_1}{2^n},
\] 
where each $a_i$ is either $0$ or~$1$.
Note that this is equivalent to $a/2^n$, where
\[
a=a_na_{n-1}\ldots a_1=a_n2^{n-1}+a_{n-1}2^{n-2}+\cdots+a_12^0.
\]
Then if $a$ is an $n$-bit integer with $n>k$, we have
\[
\begin{array}{ccl}
e(a/2^k)&=&e(a_n2^{n-k-1}+a_{n-1}2^{n-k-2}+\cdots+a_{k+1}+a_k/2+\cdots+a_1/2^k)\\
&=&e(a_k/2+a_{k-1}/2^2+\cdots+a_1/2^k)\\
&=&e(0.a_ka_{k-1}\ldots a_1).\\
\end{array}
\]

The quantum Fourier transform on an $N$-dimensional space evolves the basis state vector $\ket a$ to the state
\begin{equation}
\label{eq:QFT}
\QFT_N:\ket a\mapsto\frac{1}{\sqrt N}\sum_{j=0}^{N-1}e^{2\pi iaj/N}\ket j,
\end{equation}
where the orthonormal basis vectors of the space are enumerated $\ket0$, $\ket1$, $\ldots$, $\ket{N-1}$.
If $N$ is composite, say $N=MK$ for some integers $M$ and $K$, then we can break up the quantum Fourier transform into the tensor product of at least two distinct unentangled parts.
By the division algorithm, we can let $j=mK+k$, where $0\le m<M$, and $0\le k<K$.
We assume that we are using the positional notation with radix $K$, so that $\ket j=\ket m\otimes\ket k$.
After substituting for $j$ and $N$, the sum in Equation~\ref{eq:QFT} becomes
\begin{equation}
\label{eq:disQFT}
\begin{split}
\QFT_N\ket a&=\frac{1}{\sqrt{MK}}\sum_{m=0}^{M-1}\sum_{k=0}^{K-1}e^{2\pi ia(mK+k)/MK}\ket m\otimes\ket k\\
&=\frac{1}{\sqrt{MK}}\sum_{m=0}^{M-1}\sum_{k=0}^{K-1}e^{2\pi iam/M}e^{2\pi iak/N}\ket m\otimes\ket k\\
&=\biggl(\frac{1}{\sqrt M}\sum_{m=0}^{M-1}e^{2\pi iam/M}\ket m\biggr)\otimes\biggl(\frac{1}{\sqrt K}\sum_{k=0}^{K-1}e^{2\pi iak/N}\ket k\biggr)\\
&=\bigl(\QFT_M\ket a\bigr)\otimes\biggl(\frac{1}{\sqrt K}\sum_{k=0}^{K-1}e^{2\pi iak/N}\ket k\biggr).\\
\end{split}
\end{equation}
Equation~\ref{eq:disQFT} deserves some commentary, in that we seem to be confusing mathematical notation for actual physical reality.
The content of this equation is that we can take the $N$-dimensional space $A_N$ and decompose it as a tensor product $A_M\otimes A_K$.
We then get the $\QFT$ of \ket{a\bmod M} in $A_M$, and tensor that with the result of the operation specified in $A_K$, with the final result of the computation being the $\QFT$ of $\ket a$ in $A_N$.
Hopefully this sloppiness will not cause to much confusion to the reader, as it is a mathematical artifice too show that the quantum Fourier transform is not entangled. 
Now we assume that $N=2^n$ is a power of 2.
We let $K=2$, and $M=2^{n-1}$, and then substitute $M$ and $K$ into the Equation~\ref{eq:disQFT} above, finding
the equation
\begin{equation}
\label{eq:disQFT2}
\QFT_{2^n}\ket a=\bigl(\QFT_{2^{n-1}}\ket a\bigr)\otimes\oneover\bigl(\ket0+e(a/2^n)\ket1\bigr).
\end{equation}

We continue inductively, eventually arriving at:
\begin{equation}
\begin{split}
\QFT_{2^n}\ket a&=\frac{1}{\sqrt{2^n}}\bigl(\ket0+e(a/2)\ket1\bigr)
\otimes\bigl(\ket0+e(a/2^2)\ket1\bigr)
\otimes\cdots\\
&\quad\otimes\bigl(\ket0+e(a/2^n)\ket1\bigr).
\end{split}
\end{equation}
This formula allows an implementation of the $n$-bit quantum Fourier transform using a total of $n$ Hadamard gates, $n(n-1)/2$ controlled rotation gates, and $\lfloor n/2\rfloor$ swap gates, as shown in Coppersmith's paper~\cite{bib:Coppersmith}.

\section*{The Approximate QFT}
The approximate quantum Fourier transform (which I call $\AQFT$ throughout this paper) is implemented by deleting all controlled rotation gates with a tolerance finer than a certain threshold.
If the threshold is $k$, then the approximate QFT on $n$ qubits is given by
\begin{equation}
\begin{split}
&\AQFT\ket a=\frac{1}{\sqrt{2^n}}\bigl(\ket0+e(a/2)\ket1\bigr)\otimes\bigl(\ket0+e(a/2^2)\ket1\bigr) \\
&\quad\otimes\cdots\otimes\bigl(\ket0+e(a/2^k)\ket1\bigr)\otimes\bigl(\ket0+e(0.a_{k+1}a_k\ldots a_2)\ket1\bigr)\\
&\quad\otimes\bigl(\ket0+e(0.a_{k+2}a_k\ldots a_3)\ket1\bigr)\otimes\cdots\\
&\quad\otimes\bigl(\ket0+e(0.a_na_{n-1}\ldots a_{n-k+1})\ket1\bigr).\\
\end{split}
\end{equation}
Written this way, the approximate QFT can be put into a useful form for analysis.
First, we need to introduce some new notation.
Define $\Tr(k;m;a)$, rep\-resenting {\sl truncation}, as being a function that takes the number
\[
a=a_n2^{n-1}+a_{n-1}2^{n-2}+\cdots+a_12^0
\]
and nonnegative integers $k$ and $m$, where $m\le n$, and yields the number
\[
\Tr(k;m;a)=0.\underbrace{00\ldots0}_{\mbox{$k$ 0's}}a_ma_{m-1}\ldots a_1,
\]
so that this is the $m$ least significant digits of $a$ multiplied by $2^{-(m+k)}$.
Note that 
\begin{equation}
\label{eq:TruncIneq}
0\le\Tr(k;m;a)+\Tr(k;m;b)-\Tr(k;m;a+b)\le2^{-k},
\end{equation}
which captures the fact that carry bits propogate through the addition.
The error is bounded above because the only possible non-zero difference between the sum of the truncations and the truncation of the sum is the number
\[
0.\underbrace{000000\ldots0}_{\mbox{$k-1$ 0's}}1=2^{-k}.
\]
From~(\ref{eq:TruncIneq}), it is easy to make a inductive argument to prove that
\begin{equation}
\label{eq:TruncSumIneq}
0\le\biggl(\sum_{j=1}^n\Tr(k;m;a_j)\biggr)-\myTr_k^m\biggl(\sum_{j=1}^na_j\biggr)\le(n-1)\cdot2^{-k}.
\end{equation}
Also, it is easy to see that
\[
\begin{array}{ll}
e(a/2^m)&=e(0.a_ma_{m-1}\ldots a_1)\\
&=e(0.a_ma_{m-1}\ldots a_{m-k+1}+\Tr(k;m-k;a)).\\
\end{array}
\]
We can therefore write the approximate quantum Fourier transform as
\begin{equation}
\label{eq:completeAQFT}
\begin{split}
&\AQFT\ket a=\frac{1}{\sqrt{2^n}}\bigl(\ket0+e(a/2)\ket1\bigr)\otimes\bigl(\ket0+e(a/2^2)\ket1\bigr)\\
&\quad\otimes\cdots\otimes\bigl(\ket0+e(a/2^k)\ket1\bigr)\otimes\bigl(\ket0+e(a/2^{k+1}-\Tr(k;1;a))\ket1\bigr)\\
&\quad\otimes\bigl(\ket0+e(a/2^{k+2}-\Tr(k;2;a))\ket1\bigr)\otimes\cdots\\
&\quad\otimes\bigl(\ket0+e(a/2^n-\Tr(k;{n-k};a))\ket1\bigr).\\
\end{split}
\end{equation}

\section*{Rotation by $b$}
We have seen how to compute the approximate quantum Fourier transform of a state vector $\ket a$. 
At this point, we want to see the action of controlled rotations conditioned on bits from $b$.
If we take the last qubit in Equation~\ref{eq:completeAQFT}, we see that after the application of the approximate QFT, its unnormalized state is
\[
\ket{\phi_n}=\ket0+e(0.a_na_{n-1}\ldots a_{n-k+1})\ket1
\]
and is rotated to the state
\[
\Psi_b\ket{\phi_n}=\ket0+e(0.a_na_{n-1}\ldots a_{n-k+1}+0.b_nb_{n-1}\ldots b_{n-k+1})\ket1
\]
by the action of $\Psi_b$~\cite{bib:Draper}.
This state is equivalent to 
\[
\Psi_b\ket{\phi_n}=\ket0+e\bigl((a+b)/2^n-\Tr(k;{n-k};a)-\Tr(k;{n-k};b)\bigr)\ket1.
\]
Note that the correct state, {\it i.e.}, if we were to perform an approximate QFT on the vector $\ket{a+b}$, is 
\[
\ket{\psi_n}=\ket0+e\bigl((a+b)/2^n-\Tr(k;{n-k};a+b)\bigr)\ket1.
\]
Similar statements hold for the other qubits.
We discover the error introduced by using the approximate transform by rewriting this state as
\begin{equation*}
\begin{split}
\Psi_b\ket{\phi_n}&=\ket{\psi_n}+\Bigl\lbrace e((a+b)/2^n)\bigl\lbrack e\bigl(-\Tr(k;{n-k};a)-\Tr(k;{n-k};b)\bigr)\\
&\quad\quad-e\bigl(-\Tr(k;{n-k};a+b)\bigr)\bigr\rbrack\ket1\Bigr\rbrace.\\
\end{split}
\end{equation*}
Then $\ket{\psi_n}$ will be evolved back to the correct answer of $a+b$ by the application of the inverse approximate QFT, $\AQFT^\dag$.
The remainder of the state vector enclosed in braces is an error term.
There is also a normalization factor of $1/\sqrt2$ that I have not included.

After the application of $\Psi_b$ and the approximate QFT, the $a$ register has been evolved to a state
\begin{equation}
\label{eq:ErrorEx}
\begin{split}
&\Psi_b\AQFT\ket a=\ket{\psi_1}\otimes\ket{\psi_2}\otimes\cdots\otimes\ket{\psi_k}\otimes(\ket{\psi_{k+1}}+\ket{\gamma_{k+1}})\\
&\qquad\otimes(\ket{\psi_{k+2}}+\ket{\gamma_{k+2}})\otimes\cdots\otimes(\ket{\psi_n}+\ket{\gamma_n}),\\
\end{split}
\end{equation}
where $\ket{\gamma_j}$ represents the error on qubit $j$.
If we expand this expression out, we have $2^{n-k}$ terms.
There will be $\binom{n-k}{m}$ terms with $m$ errors for each $0\le m\le n-k$.
We have the identity that
\[
e^{i\theta}-e^{i\alpha}=\int_\alpha^\theta ie^{it}\,dt.
\]
So
\begin{equation}
\begin{split}
|e^{i\theta}-e^{i\alpha}|&=\left|\int_\alpha^\theta ie^{it}\,dt\right|\\
&\le\left|\int_\alpha^\theta|ie^{it}|\,dt\right|\\
&=|\theta-\alpha|\\
\end{split}
\end{equation}
for any $\theta$ and $\alpha$.
Combining this with the inequality (\ref{eq:TruncIneq}) yields that, for every $j$ such that $n-k<j\le n$,
\[
\|\ket{\gamma_j}\|\le\pi\sqrt2\cdot2^{-k}.
\]
We also know that
\[
\|\ket a\otimes\ket b\|=\|\ket a\|\cdot\|\ket b\|,
\]
and that $\|\ket{\psi_i}\|=1$ for every $1\le i\le n$; therefore, the effect of having $m$~errors in a term is that the magnitude of the error on that term is equal to the product of the magnitudes of each of the $m$~errors.
Since $\AQFT$ is unitary, the magnitude of the error vector is preserved when we apply $\AQFT^\dag$.
Using the triangle inequality ($\|\sum_ix_i\|\le\sum_i\|x_i\|$) to bound the error in the state given in Equation~\ref{eq:ErrorEx}, we see that the magnitude of the error must be bounded above by
\begin{equation*}
\sum_{m=1}^{n-k}\binom{n-k}{m}(\pi\sqrt2\cdot2^{-k})^m=(1+\pi\sqrt2\cdot2^{-k})^{n-k}-1
\end{equation*}
The probability of error is this number squared.
When we do a first-order expansion of this expression, we get
\begin{equation}
\label{eq:BoundEstimate}
\pi(n-k)\cdot2^{-k}\cdot\sqrt2
\end{equation}
as a rough estimate for the bound.
Thus $k$ does not have to be very large compared to $n$ to maintain a very small probability of error in the computation.
For example, if $n=1000$, and $k=30$, then the bound given in (\ref{eq:BoundEstimate}) is approximately $4.0\times10^{-6}$; the probability of getting an error is smaller than $1.6\times10^{-11}$.
One can use~(\ref{eq:TruncSumIneq}) to prove that if we perform $m$ $k$-truncated additions---i.e., add $m+1$ numbers---the magnitude of the error vector is bounded above by
\[
(1+\pi\sqrt2\cdot2^{-k}\cdot m)^{n-k}-1,
\]
which also goes quickly to zero, and thus implies that there is a high probability of getting the correct answer when one uses a sufficient number of controlled rotation gates.

\section*{Summary}
We have derived a bound on the error introduced by truncating in Draper's quantum addition algorithm.
This bound goes quickly toward zero, and therefore promises that only a moderate number of rotation gates are required to ensure reliable addition if we use Draper's algorithm in connection with reliable quantum computational gates.


\end{document}